# MODE-LOCKED PULSE OSCILLATION OF A SELF-RESONATING ENHANCEMENT OPTICAL CAVITY

Y. Hosaka*, QST, Takasaki, Gunma 370-1292, Japan
Y. Honda, T. Omori, J. Urakawa, KEK, Tsukuba, Ibaraki 305-0801, Japan
A. Kosuge, University of Tokyo, Kashiwa, Chiba 277-8581, Japan
K. Sakaue, University of Tokyo, Bunkyo, Tokyo 113-8656, Japan
T. Takahashi, Hiroshima University, Higashi-Hiroshima, Hiroshima 739-8530, Japan
Y. Uesugi, Tohoku University, Sendai, Miyagi 980-8577, Japan
M. Washio, Waseda University, Shinjuku, Tokyo 169-8555, Japan

*Abstract*

A power enhancement optical cavity is a compelling means of realizing a pulsed laser with a high peak power and high repetition frequency, which is not feasible using a simple amplifier scheme. However, a precise feedback system is necessary for maintaining the narrow resonance condition of the optical cavity; this has become a major technical issue in developing such cavities. We have developed a new approach that does not require any active feedback system, by placing the cavity in the outer loop of a laser amplifier. We report on the first demonstration of a mode-locked pulse oscillation using the new system.

## INTRODUCTION

A power enhancement optical cavity is a compelling means of realizing a pulsed laser with a high peak power and high repetition frequency. In this approach, a laser pulse train is injected into an external empty optical cavity whose resonance frequency matches the laser pulse frequency. By stacking many laser pulses coherently, a high-peak-power laser pulse can be realized in the cavity. Since this approach does not require an additional amplifier that limits the power density, it becomes possible to reach much higher power than that provided by the amplifier.

We have been studying the X-ray and γ-ray production via laser-Compton scattering [1-3] using the high-peak-power laser pulse in the enhancement cavity. In this application, the photon flux is directly related to the stored laser power in the cavity. Thus, it is necessary to increase the power enhancement factor of the cavity.

Since the enhancement factor depends on the round-trip loss of the cavity, high-reflectance cavity mirrors are required. An enhancement factor of $\sim 10^6$ can be realized when multi-layered dielectric mirrors with a loss on the order of parts per million [4] are used, however, the high enhancement factor has a drawback from the perspective of cavity length control. For example, the required accuracy for controlling such a high-enhancement cavity is on the order of sub-picometer. This can be realized using a state-of-the-art feedback system [5-7] in an environment that shuts out mechanical and electrical disturbances. Although maintaining such a sharp cavity resonance is not impossible, it is technically quite difficult.

* hosaka.yuji@qst.go.jp

In order to eliminate the technical difficulties associated with controlling the cavity resonance, we have proposed a new approach referred to as a self-resonating enhancement cavity scheme [8]. The scheme is an integrated system of a laser amplifier and an enhancement optical cavity. In this scheme, the system automatically provides selective amplification of the laser light that resonates in the cavity. Hence, active controlling is not required to maintain the cavity resonance.

A demonstration of the self-resonating cavity operating in continuous-wave (CW) mode was provided in our previous work [8], where a cavity with an effective finesse of 394,000 was kept resonating without any feedback control. Now, we have been developing pulse oscillation of the self-resonating enhancement cavity. We provide the first demonstration of the pulse oscillation in the self-resonating enhancement cavity in this work.

## PRINCIPLES

The conceptual scheme of a self-resonating enhancement optical cavity is shown in Fig. 1. The empty enhancement cavity is incorporated into an outer optical loop, which contains a laser amplifier. The transmitted light through the cavity is amplified and then injected back into the cavity again. Because the enhancement cavity can be understood as an inner optical loop, the overall system is a double-loop oscillator. Since the light out of the resonance condition is reflected by the incident mirror of the cavity, the cavity works as a filter that transmits the light satisfying the resonance condition. Therefore, the transmitted light automatically satisfies the resonance condition in this system. The oscillation starts from a spontaneous emission of the amplifier and automatically grows to a steady state balancing the amplifier gain with the loop loss.

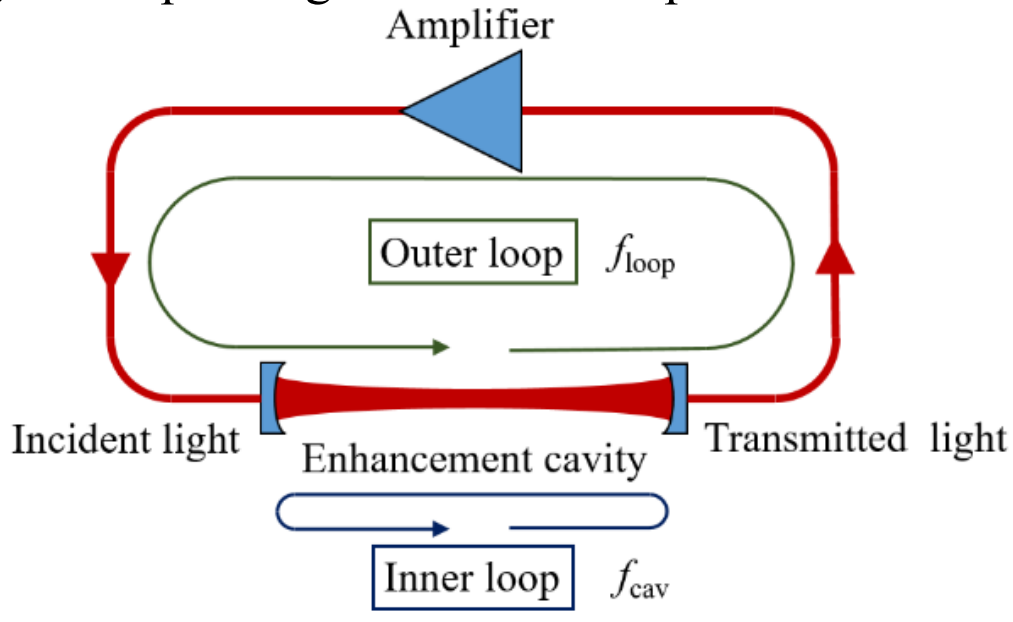

Figure 1: Self-resonating enhancement optical cavity.

Although CW self-resonating oscillation can be achieved without any controls of the cavity length, mode-locked pulse self-resonating oscillation requires a condition that is not so difficult. There are two characteristic repetition frequencies in the system: the frequency of the outer loop ($f_{\text{loop}}$) and the frequency of the enhancement cavity ($f_{\text{cav}}$) (Fig. 1). In order to obtain pulse oscillation in the system, the following relationship should be satisfied:

$$f_{\text{cav}} = n f_{\text{loop}}, \quad (1)$$

where n is an integer that represents the number of circulating pulses in the outer loop. This equation represents the condition for a pulse that travelled around the outer loop to overlap a reciprocating pulse in the cavity.

The equation should be satisfied with the accuracy according to the pulse width. Since the mode-locked pulse width in this study was several picoseconds, the loop length should be adjusted with the order of 100 μm. It is important to note that this requirement is more easily achievable than the required accuracy in conventional high-finesse cavities, which is the order of sub-picometer.

## EXPERIMENTAL PROCEDURES

Since the pulse energy is important for mode-locked oscillation, we adopted a long enhancement cavity in this study to obtain a higher pulse energy with a lower repetition frequency. The constructed Fabry-Perot-type optical cavity was folded into a V-shape in order to achieve a long cavity length in a limited space. The cavity length was 3.62 m, which corresponds to a repetition frequency of 41.4 MHz. The mirrors at both ends were concave; each had a radius of curvature of 2.0 m and a reflectance of 98.6%. The folding mirror at the center of the cavity was a planar mirror with a reflectance of 99.9%. The finesse of the cavity was calculated to be 220 at 1030 nm according to the specifications of the mirrors, and was measured to be 170 at 1064 nm by using an external laser. One of the end mirrors was mounted on a piezo-controlled movable stage, which allowed the cavity length to be scanned.

The schematic of the whole system is shown in Fig. 2. Yb-doped single-mode fiber with a length of approximately 1.0 m was used as a laser amplifier. Pumping was performed by a laser diode (LD) with a wavelength of 976 nm through a wavelength division multiplexed coupler (WDM). The output of the fiber amplifier was emitted to free space through a collimator. A half wave plate (HWP), quarter wave plates (QWP), and polarizing beam splitter (PBS) were installed for mode-locking. The reflected light of the PBS was used for monitoring the oscillation, and the isolator was placed to determine the propagation direction. Two beam samplers (BS) were inserted at positions upstream and downstream from the cavity to measure the incident, reflected, and transmitted power of the cavity. An optical band-pass filter (BPF) with a center wavelength of 1030 nm and a transmission width of 3 nm (FWHM) was installed for stabilizing the mode-locked pulse formation. The laser was then injected into a fiber section with another collimator. An optical delay line was installed in the fiber section to adjust the outer loop frequency according to Eq. (1). Both $f_{\text{loop}}$ and $f_{\text{cav}}$ was measured precisely and individually in advance. All the optical fibers were single mode and connected using fusion splicing.

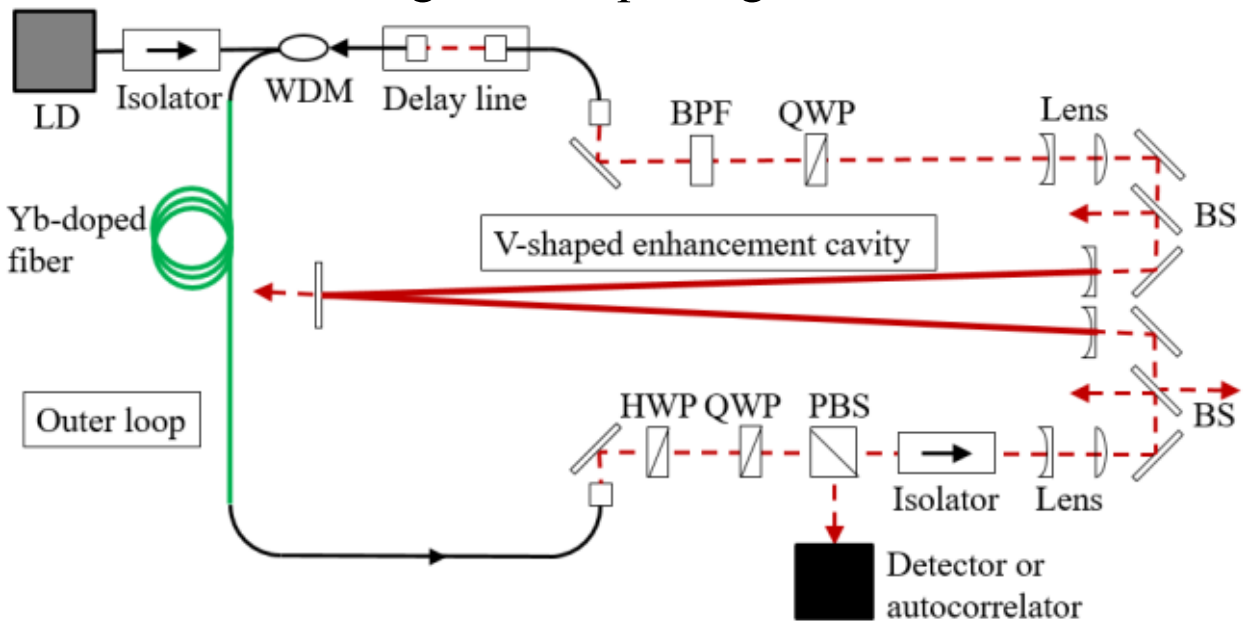

Figure 2: Schematic of the pulsed fiber laser oscillator with the self-resonating enhancement cavity.

## RESULTS

The pulse oscillation was realized in the self-resonating cavity scheme. The pulse width was measured using an autocorrelation method. Figure 3 shows a measured autocorrelation trace and the pulse width (FWHM) was estimated to be 2–4 ps.

By adjusting the angle of the wave plates in the outer loop, the oscillation could be switched to CW operation. We measured the emission spectrum for the CW and pulse oscillation using an optical spectrum analyser (Fig. 4). The spectrum width in the pulse oscillation was measured to be approximately 6 nm (FWHM) and much broader than in the case of CW operation. This result confirms the mode-locked pulse oscillation in self-resonating optical cavity.

In order to check the tolerance of the repetition frequency adjustment, we varied $f_{\text{loop}}$ using the delay line, and observed whether or not the pulse oscillation was realized. The range capable of realizing the pulse oscillation was determined to be ±0.2 mm.

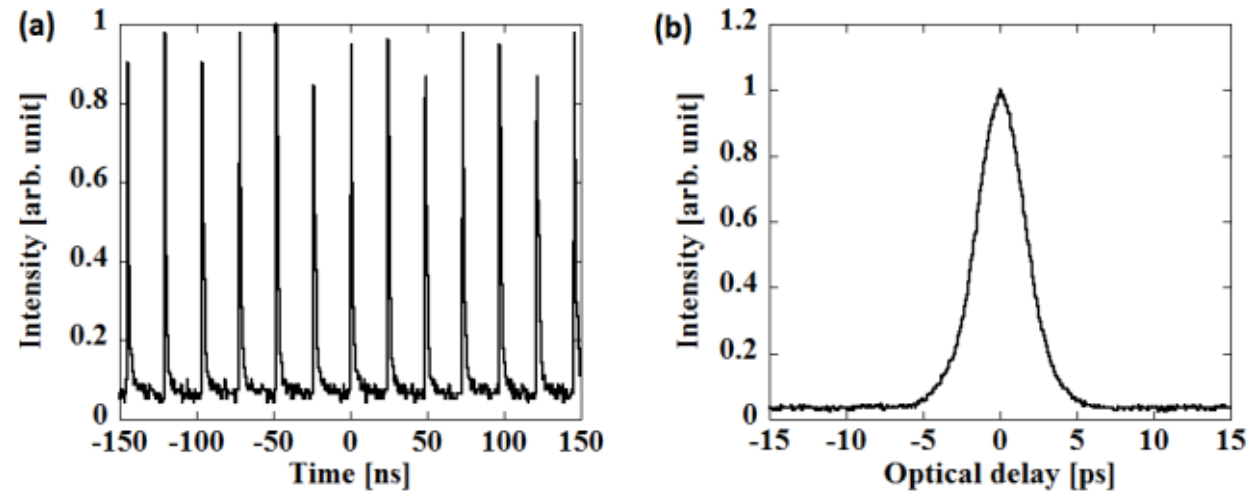

Figure 3: (a) Temporal waveform and (b) autocorrelation trace of pulse oscillation in the self-resonating cavity.

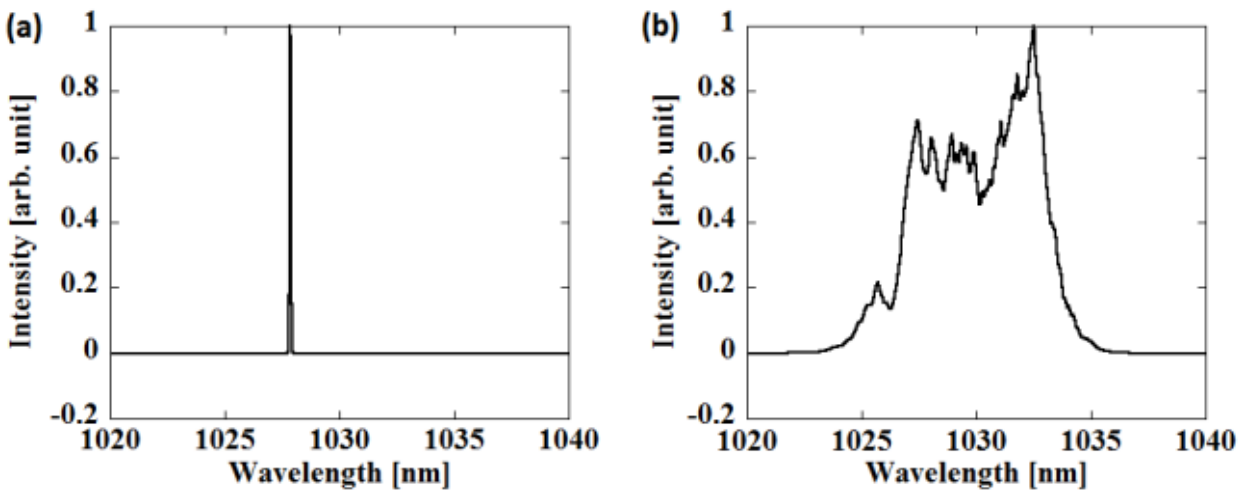

Figure 4: Wavelength spectra of (a) CW oscillation and (b) mode-locked pulse oscillation in the self-resonating cavity.

Table 1: Laser Power of the Self-Resonating Cavity

| Operation mode | Incident power | Reflected power | Transmitted power | Matching efficiency | Stored power |
|---|---|---|---|---|---|
| Mode-locked | 111 mW | 31 mW | 47 mW | 0.72 | 4200 mW |
| CW | 164 mW | 24 mW | 82 mW | 0.85 | 7700 mW |

By measuring laser powers at monitoring ports, we determined the incident, reflected, and transmitted powers of the cavity. The power stored in the cavity was also estimated from the transmitted power of the planar mirror in the cavity center. The results are summarized in Table 1. The stored power was estimated to be 4,200 mW, whereas the incident power was 111 mW in the mode-locked pulse oscillation. This corresponds to an enhancement factor of 38 and a matching efficiency of 0.72. We also performed the measurement in the CW oscillation without changing the optical alignment or amplifier gain. The enhancement factor was found to be 47, and the matching efficiency was found to be 0.85, hence, the matching efficiency in the pulse oscillation was lower than that in the CW oscillation.

## DISCUSSION

The measured pulse width was found to be 2–4 ps in the self-resonating system. The transform limit, which is a theoretical minimum pulse width calculated from the wavelength width, was approximately 0.3 ps. This was shorter than the measured value; however, this difference is typical for fiber laser using an optical BPF. Thus, the pulse width of the output laser can be compressed to a value comparable to the transform limit by an external pulse compressor.

The required accuracy of the frequency condition of the scheme (Eq. (1)) was found to be ±0.2 mm for the cavity length, which corresponds to an $f_{\mathrm{loop}}$ of ±0.29 kHz and a propagation time of ±0.7 ps. The requirement should be stricter than the mode-locked pulse width (2–4 ps) because a pulse that travelled around the outer loop should overlap a reciprocating pulse in the cavity. Since the resonance width of the optical cavity is approximately 2.3 nm, it is worth noting that the resonance condition was automatically satisfied in the self-resonating scheme.

We confirmed power enhancement in the self-resonating cavity. The achieved pulse energy and peak power were calculated to be approximately 100 nJ and 30 kW, respectively. Although this experiment was performed for the initial demonstration using a low-finesse cavity, it can be expected that a pulse energy on the order of mJ using a high-finesse cavity is achievable as well as CW self-resonating oscillation [8]. We also performed experiments with two different settings of the cavity mirrors: namely 50% and 90% reflectance. The required accuracy of the path length adjustment was found to be comparable to the case with 98.6% reflectance mirrors. So far, we have not faced difficulties in increasing the cavity finesse while using these low-finesse cavities. We plan to test a pulsed self-resonating cavity with a higher finesse in subsequent studies.

On the other hand, the stability of the obtained mode-locked pulse oscillation was insufficient for practical applications. The pulse oscillation maintained for only several seconds and restarted with a different pulse width. We therefore plan to investigate methods for stabilizing the pulse oscillation in the self-resonating cavities.

## CONCLUSION

The self-resonating enhancement optical cavity is an attractive scheme to realize the resonance condition of an optical cavity without any active feedback system. In this work, we investigated the possibility of realizing a pulse oscillation in the scheme. We succeeded in demonstrating a mode-locked pulse oscillation in the self-resonating enhancement cavity. We note that the required accuracy for adjustment of the path length in this scheme is rather coarse, and hence is much more easily achieved than that required for resonating in a conventional scheme. We also confirmed that power enhancement occurred in the cavity. For the use of the self-resonating cavity in practical applications, we plan to develop a pulsed self-resonating cavity with higher finesse and more stability in subsequent studies.

## ACKNOWLEDGEMENTS

This work was supported by JSPS KAKENHI Grant Number JP25246039 and JP18K18308.